\documentclass[aps,twocolumn,prc,superscriptaddress,showpacs,nofootinbib,floatfix,amssymb,amsfonts,amsmath]{revtex4-2}

\usepackage{graphicx}% Include figure files
\usepackage{dcolumn}% Align table columns on decimal point
\usepackage{bm}% bold math
\usepackage{xcolor}
\usepackage{amsmath}    % need for subequations
\usepackage{amsfonts}   %note how statements can be commented out
\usepackage{amssymb}
\usepackage{graphicx}   % for figuresManMan//

\begin{document}

\title{Differential charge radii: self-consistency and proton-neutron interaction effects }

\author{U. C. Perera}
\affiliation{Department of Physics and Astronomy, Mississippi
State University, MS 39762}

\author{A. V. Afanasjev}
\affiliation{Department of Physics and Astronomy, Mississippi
State University, MS 39762}

\date{\today}

\begin{abstract}
 
The analysis of self-consistency and proton-neutron interaction
effects in the buildup of differential charge radii  has been carried out
in covariant density functional theoretical calculations without pairing
interaction. 
Two configurations of the $^{218}$Pb nucleus, generated by 
the occupation of the neutron $1i_{11/2}$ and  $2g_{9/2}$ subshells, are
compared with the ground state configuration in $^{208}$Pb.
The interaction 
of added neutron(s) and the protons forming the $Z=82$ proton core is
responsible for a major contribution to the buildup of differential charge radii.
It depends on the overlaps of  proton and neutron wave functions and 
leads to a redistribution  of single-particle density of occupied proton 
states which in turn modifies the charge radii. Self-consistency effects affecting 
the shape of proton potential, total proton densities and the energies of the 
single-particle proton states provide only secondary contribution
to differential charge radii.  The buildup of differential charge radii is a combination
of single-particle and collective phenomena.  The former is due to proton-neutron
interaction, the impact of which is state dependent, and the latter reflects
the fact that all occupied proton single-particle states contribute to this process.
The neglect of either one of these aspects of the process by ignoring 
proton-neutron interaction and self-consistency effects as it is done in 
macroscopic+microscopic approach or by introducing the core as in spherical
shell model introduces uncontrollable errors and restricts the applicability
of such approaches to the description of differential charge radii.
The analysis also indicates that
both the Coriolis interaction in odd and odd-odd deformed nuclei and the residual
interaction between unpaired proton and neutron in odd-odd nuclei
could affect the odd-even staggering in charge radii if their impact
on the wave function of the ground state of these nuclei is appreciable.
\end{abstract}
\maketitle

%%%%%%%%%%%%%%%%%%%%
\section{Introduction}
%%%%%%%%%%%%%%%%%%%%

Charge radii are among the most fundamental properties of atomic nuclei and during 
the last decade there was a significant increase in experimental and theoretical studies 
of this physical observable. The experimental results were reviewed in Refs.\ 
\cite{AM.13,CMP.16} and recent experimental investigations were summarized in the 
introduction of Ref.\ \cite{PAR.21}. The introduction to the latter publication provides 
also the overview of theoretical efforts.  Theoretical calculations within different density 
functional theories (DFTs) provide a quite accurate global description of experimental
charge radii presented in the compilation of Ref.\ \cite{AM.13}: the rms deviations
of calculated charge radii $r_{ch}$ from experimental ones are at the level of 
$\approx 0.03$ fm \cite{AARR.14} which corresponds to high average precision 
of 0.625\% in the prediction of charge radii (see Ref.\ \cite{PAR.21}).
 
  The changes of the charge radii within the isotopic chain are measured with high 
precision using laser spectroscopy (see Refs.\ \cite{AM.13,CMP.16}). Thus, the 
differential mean-square (ms) charge radii (see Eq.\ (\ref{diff-radii}) below for definition), 
measured with high precision within the isotopic chains, become an important quantity.
They have been studied within {\it ab initio} approaches (see, for example, Refs.\ 
\cite{Ca-radii.2016,K-radii.21}),  non-relativistic DFTs based on zero range 
Skyrme forces (see Refs.\ \cite{TBFHW.93,RF.95,Gorges-Sn-radii.19,NOSW.22}),
Fayans functional (see Refs.\ \cite{FTTZ.94,FTTZ.00,RN.17}) and finite range Gogny forces 
(see Refs.\ \cite{RSRP.10,Ca-radii.2016}), non-relativistic Hartree-Fock-Bogoluibov (HFB) 
approach with a  finite-range Yukawa interaction \cite{Nakada.15,NI.15,Nakada.19} and
covariant density functional  theory (CDFT) (see Refs.\ 
\cite{SLR.93,Sharma1995_PRL74-3744,Pb-Hg-charge-radii-PRL.21,PAR.21,NOSW.22}). 

   Different aspects of the buildup of  differential charge radii within the isotopic
chain with increasing neutron number have been investigated in these papers. In
particular, it was found that the evolution of the charge radii within the isotopic chain 
with increasing neutron number is defined by the pull on the proton states generated 
by neutrons gradually added to the nuclear system \cite{RF.95,GSR.13,Pb-Hg-radii-PRC.21,PAR.21}.
The most investigated case here is the kink in charge radii at $N=126$ and the evolution
of charge radii above $N=126$ in the Pb isotopic chain. The pattern of these effects
critically depends on the occupation of the $2g_{9/2}$ and $1i_{11/2}$ orbitals, on
their relative energies, and on how close they are in energy \cite{RF.95,GSR.13,PAR.21}:
the kink is generated only when neutron $1i_{11/2}$ orbitals are substantially occupied
in the nuclei with $N>126$.

 Ref.\ \cite{GSR.13} provided a hint on microscopic origin of 
this difference by revealing  that it is traced back to the nodal structure of these two 
orbitals ($n=1$ for $1i_{11/2}$ and $n=2$ for  $2g_{9/2}$, where $n$ stands for principal 
quantum number) and the overlap of their wavefunctions with those of the proton states. 
However, as follows from the present study the interpretation of the pull of these neutron 
states on proton orbitals via the symmetry energy does not corresponds to real physical 
situation since it is related to proton-neutron interaction. The detailed global analysis 
of the impact of the occupation of neutron single-particle orbitals in the vicinity of 
spherical neutron shell closures generalized the results of Ref.\ \cite{GSR.13} to whole
nuclear chart (see discussion of Fig. 32 in Ref.\ \cite{PAR.21}). It revealed  strong 
correlations between the principal quantum number $n$ 
of the single neutron orbital occupied above the neutron shell closure and the impact 
of the occupation of this orbital on differential charge radii: in a given isotopic chain the
largest impact on differential charge radii is provided by the occupation of the neutron
orbital with the lowest $n$.  As a consequence, a significant occupation of the
$n=1$ neutron subshell above the neutron shell closure is required for a creation of
the kink in differential charge radii at this closure (see Ref.\ \cite{PAR.21}).

    Despite all these studies, there are the aspects of the process of the buildup
of differential charge radii which are not completely understood on microscopic
level and which have not been discussed in the literature. Thus, the goal of the 
present paper is to fill these gaps in our knowledge and to perform detailed 
studies of the impact of self-consistency effects and the interaction between 
neutron(s) added to a reference nucleus and the protons forming the proton 
subsystem on the buildup of differential charge radii. Of particular interest is 
the balance of these two types of contributions and the microscopic mechanisms 
affecting the changes of charge radii in the isotopic chain with increasing neutron
number. The results  of these studies will also allow to answer the question of the
applicability of different theoretical frameworks to the description of differential
charge radii and potential contributing factors to odd-even staggering in
charge radii.

      The paper is organized as follows. Sec.\ \ref{Theory} provides a brief outline 
of theoretical formalism and the discussion of physical observables under study. The 
impact of self-consistency and proton-neutron interaction effects on differential charge 
radii is discussed in Sec.\ \ref{self-const}. Sec.\ \ref{gen-observation} is dedicated to 
general observations following from this study.  Finally, Sec.\ \ref{Concl} summarizes 
the results of our paper.

%%%%%%%%%%%%%%%%%%%%%%%%%%%%%%
\section{Theoretical formalism and physical observables} 
\label{Theory}
%%%%%%%%%%%%%%%%%%%%%%%%%%%%%%

    Theoretical calculations have been performed within the framework of covariant density 
functional theory (CDFT) \cite{VALR.05} employing the modified version of the computer 
code restricted to spherical  symmetry used in Ref.\ \cite{AF.05-dep}.    Since the details of the 
CDFT framework are widely available (see, for example, Ref.\  \cite{VALR.05}), we focus on the 
physical quantities of the interest. The pairing correlations are neglected in the calculations in order 
to better understand the underlying physical mechanisms.  The calculations are performed with 
the NL3* covariant energy density functional (CEDF) \cite{NL3*}. Its global performance in the
description of the masses and charge radii  is well documented (see Refs.\ \cite{AARR.14,PAR.21}).
It was also recently used in the study of bubble nuclei (see Ref.\ \cite{PA.22}) the results of which 
have substantial overlap with some aspects of the present study; this is one of main reasons
for the selection of this functional. Note that it is was verified that main conclusions obtained  in 
the present paper do not depend on the selection of the functional.

 In order to better apprehend the role of self-consistency effects on differential 
charge radii we consider  ground state configuration in $^{208}$Pb and  two configurations of 
the $^{218}$Pb nucleus labeled below as "Conf-$1i_{11/2}$" and "Conf-$2g_{9/2}$".  In 
these configurations of  $^{218}$Pb, ten  neutrons outside the $^{208}$Pb core are 
located in the $1i_{11/2}$ and $2g_{9/2}$ spherical subshells, respectively. We selected 
$^{218}$Pb in order to maximize the effect of the addition of neutrons in a given spherical
subshell on proton charge radii\footnote{One can definitely consider the $1i_{11/2}$
and  $2g_{9/2}$ configurations in odd-$A$ $^{209}$Pb nucleus and this will 
completely justify the neglect of pairing which collapses because of the blocking of
odd neutron. However, this will not change the results and conclusions of the paper.}.
Note that the maximum number of neutrons which can 
be put into  the $2g_{9/2}$ and $1i_{11/2}$ spherical subshells is 10 and 12, respectively.
Thus, the selection of $^{218}$Pb nucleus corresponds to full filling of the $2g_{9/2}$ neutron
subshell and almost (two neutrons short) full filling of the $1i_{11/2}$ neutron subshell. 

    The charge radii are defined as
\begin{eqnarray}
r_{ch} = \sqrt{\left< r^2 \right>_p + 0.64}\,\,\,\, {\rm fm}
\label{r_charge}
\end{eqnarray}
where the mean square proton point radius is given by
\begin{eqnarray}
\left< r^2 \right>_p = \frac{\int r^2 \rho_{tot}^p(\vec{r}\,) d^3r} {\int \rho_{tot}^p(\vec{r}\,) d^3r} \label{rad}
\end{eqnarray}
and the factor 0.64 accounts for the finite-size effects of the proton\footnote{
   Small contributions to the charge radii originating from the electric neutron
form factor and electromagnetic spin-orbit coupling \cite{BFHN.72,NS.87} are 
neglected  in the NL3* functional (as well as in the fitting protocols of all existing 
CEDFs). More precise expressions for charge radii in CDFT are available 
but their use would require the refit of the CEDFs (see Refs.\ \cite{HP.12,Kurasawa.19} 
and discussion in Sec.\ VIII of Ref.\ \cite{PAR.21}). However, this neglect is not
critical since spin-orbit contribution to charge radii decreases with increasing 
the mass of nuclei \cite{HP.12,RN.21}  and its contribution to differential charge 
radii of the Pb isotopes is expected to be negligible \cite{RN.21}.}. 
 Then differential mean-square charge radius is given by\footnote{This quantity is 
frequently written as a function of mass number $A$. However, we prefer to define it as 
a function of neutron number $N$ since this allows to see the behavior of the
$\delta \left < r^2 \right>_p^{N,N'}$ curves at neutron shell  closures.}
\begin{eqnarray}
\delta \left < r^2 \right>_p^{N,N'} &=& \left < r^2 \right>_p(N)
                                                          - \left < r^2 \right>_p(N') = \nonumber \\
                                                      && =r^2_{ch}(N) - r^2_{ch}(N').
\label{diff-radii}
\end{eqnarray}

Note that $N'$ is the neutron number of the reference 
nucleus ($^{208}$Pb  in this paper).

   The total nucleonic density $\rho_{tot} (r)$ in a given subsystem
(proton or neutron) is built from the contributions of individual
particles as follows:
\begin{eqnarray} 
\rho_{tot}(r)  =  \sum_{i} m_i \rho^{sp}_i (r),
\end{eqnarray} 
where $m_i$ is the multiplicity of the occupation of the $i-$th subshell [$m_i = (2j_i+1)$
for a fully occupied subshell with angular momentum $j_i$] and 
$\rho^{sp}_i(r)$ is the density of the single-particle state belonging to the 
$i-$th subshell with the normalization
\begin{eqnarray}
\int \rho^{sp}_i (\vec{r}\,) d^3r = 4\pi \int r^2 \rho^{sp}_i (r) dr = 1.0 .
\label{norm-cond}
\end{eqnarray}

 Taking into account that $\int \rho_{tot}^p(\vec{r}\,) d^3r=Z$ and that all proton
subshells below the $Z=82$ shell gap are fully occupied in the proton subsystem of 
the Pb isotopes, Eq.\ (\ref{rad}) can be rewritten as
\begin{eqnarray}
\left< r^2 \right>_p = \frac{1}{Z} \sum_i (2j_i+1) \left< r^2 \right>_i^p 
\end{eqnarray} 
where 
\begin{eqnarray} 
\left< r^2 \right>_i^p = \int r^2 \rho_{i}^p(\vec{r}\,) d^3r
\end{eqnarray} 
is the proton mean square radius of the single-particle state belonging to the
$i$-th subshell. As a consequence, the differential charge radius of two
isotopes can be redefined as 
\begin{eqnarray}
\delta \left < r^2 \right>_p^{N,N'} = 
\frac{1}{Z} \sum_i (2j_i+1) [\left< r^2 \right>_i^p(N)  - \left< r^2 \right>_i^p(N')] 
\label{differ-sum} \nonumber \\
\end{eqnarray}
and its magnitude could be traced back to the modifications in 
proton mean square radius of the single-particle states generated
by the transition from the nucleus with neutron number $N'$ to the nucleus 
with $N$.  The quantity
\begin{eqnarray}
\Delta \left < r^2 \right>_i^{N,N'} = \left< r^2 \right>_i^p(N)  - \left< r^2 \right>_i^p(N')
\label{dif-sp}
\end{eqnarray} 
is denoted here as differential single-particle proton radius of the single-particle
state belonging to the $i$-th proton subshell.
 
   Note that for simplicity of the discussion of the role of the single-particle
states we consider also single-particle rms radii $r^p_i$ of the proton states 
defined as
\begin{eqnarray}
r^p_ i = \sqrt{\left< r^2 \right>_{i}^p}.
\end{eqnarray}

   The proton ($i=\pi$) and neutron ($i=\nu$) nucleonic potentials are defined 
 in the CDFT as follows:
\begin{eqnarray}
V_\pi & = & V + S + V_{Coul} , \\
V_\nu & = & V + S ,
\end{eqnarray} 
where scalar potential is given by
\begin{eqnarray} 
S(r) = g_{\sigma} \sigma(r),
\end{eqnarray}
meson defined part of the vector potential is written as
\begin{eqnarray} 
V(r) = g_{\omega} \omega_0 (r) + g_{\rho} \tau_3 \rho_0(r) ,  
\end{eqnarray}
and 
\begin{eqnarray} 
V_{Coul}(r) = e A_0(r)
\end{eqnarray}
is the Coulomb potential.

%%%%%%%%%%%%%%%%%%%%%%%%
\section{The role of self-consistency and proton-neutron interaction effects}
\label{self-const}
%%%%%%%%%%%%%%%%%%%%%%%%

%%%%%%%%%%%%%%%%%%%%%%%%%%
\subsection{Nucleonic potentials and densities}
\label{self-pot-den}
%%%%%%%%%%%%%%%%%%%%%%%%%% 
 
%%%%%%%%%%%%%%%%%%%%%%%%%%%%%%%%%%%%%%%%%%%%%%%
\begin{figure}[ht]
\centering
\includegraphics[width=8.4cm]{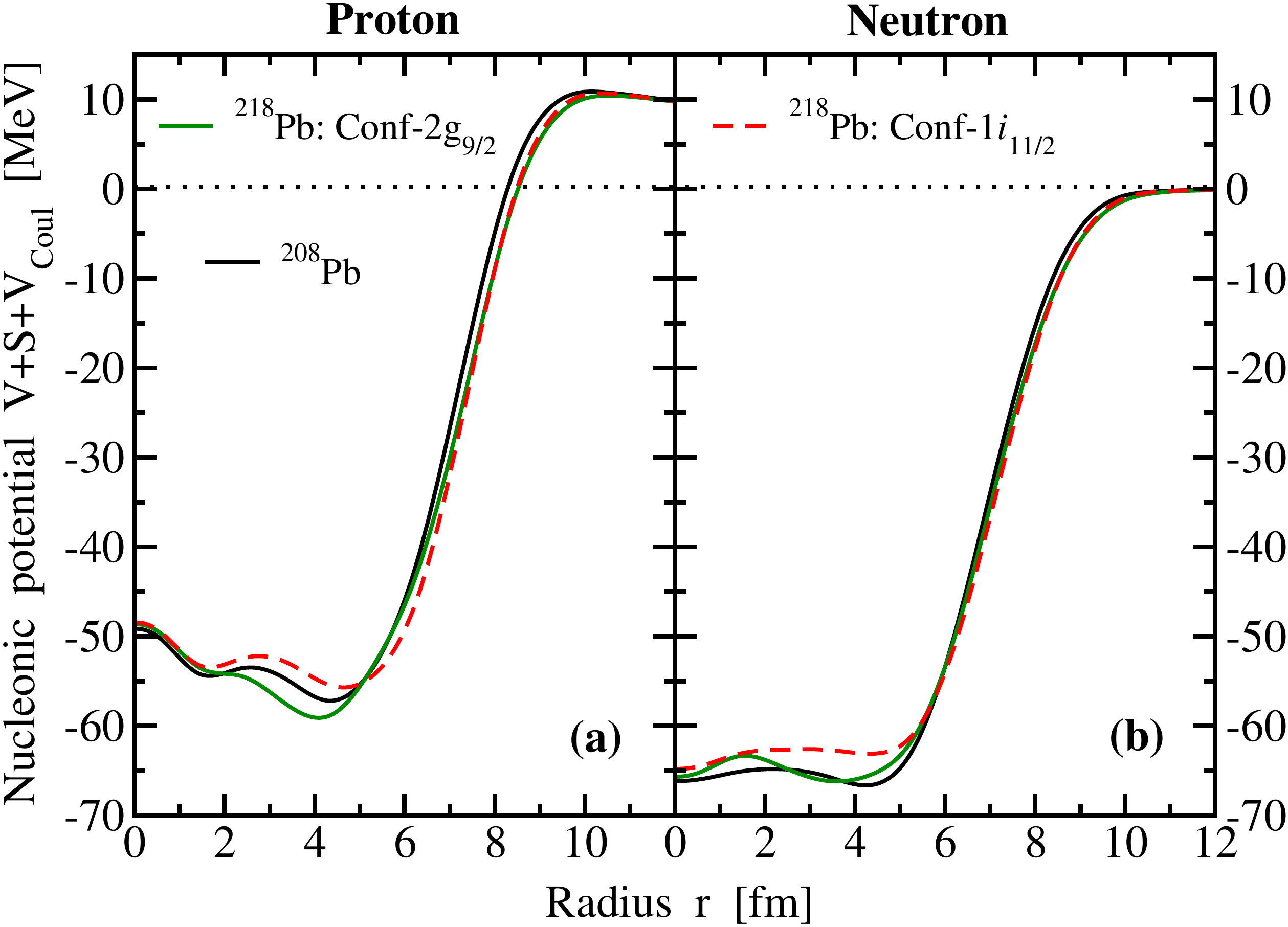}
\caption{Proton and neutron nucleonic potentials in the configurations under study. Dotted 
horizontal line shows zero energy threshold. Note that $V_{Coul}=0$ in the panel (b). 
\label{Pb_VPS_neu_pro}
}
\end{figure}
%%%%%%%%%%%%%%%%%%%%%%%%%%%%%%%%%%%%%%%%%%%%%%%%

    To better understand the impact of the occupation of different 
neutron  spherical subshells  on charge radii  of the $Z=82$ core we first consider 
how nucleonic potentials and densities change on transition from the 
$^{208}$Pb nucleus to the "Conf-$1i_{11/2}$" and "Conf-$2g_{9/2}$"
configurations in the $^{218}$Pb nucleus.

   Proton and neutron nucleonic potentials of the configurations under 
study  are shown in Fig.\ \ref{Pb_VPS_neu_pro}. The addition of ten neutrons to 
the $^{208}$Pb core only slightly increases the radius of neutron potential
and the shapes of neutron potentials of the "Conf-$1i_{11/2}$" and "Conf-$2g_{9/2}$" 
configurations in $^{218}$Pb are almost identical in the energy range between -60 
MeV and 0 MeV [see Fig.\ \ref{Pb_VPS_neu_pro}(b)]. The largest changes are 
seen at the bottom of the neutron potentials which affects mostly deep lying
neutron states. However, these potentials remain nearly flat bottom for all 
configurations of interest.

    It is interesting that the addition of ten neutrons to $^{208}$Pb triggers 
larger changes in the radial profile of proton potential than that for neutron one (compare
panels (a) and (b) in Fig.\ \ref{Pb_VPS_neu_pro}). The proton potentials  of the 
"Conf-$1i_{11/2}$" and "Conf-$2g_{9/2}$"  configurations in $^{218}$Pb are 
almost identical in the energy range between -20 MeV and 0.0 MeV  [see 
Fig.\ \ref{Pb_VPS_neu_pro}(a)]. At lower energies, the radius of the proton 
potential of the  "Conf-$1i_{11/2}$" configuration  is smaller than that of the 
"Conf-$2g_{9/2}$" one and this difference increases with decreasing energy.

%%%%%%%%%%%%%%%%%%%%%%%%%%%%%%%%%%%%%%%%%%%%%%%
\begin{figure}[ht]
\centering
\includegraphics[width=8.4cm]{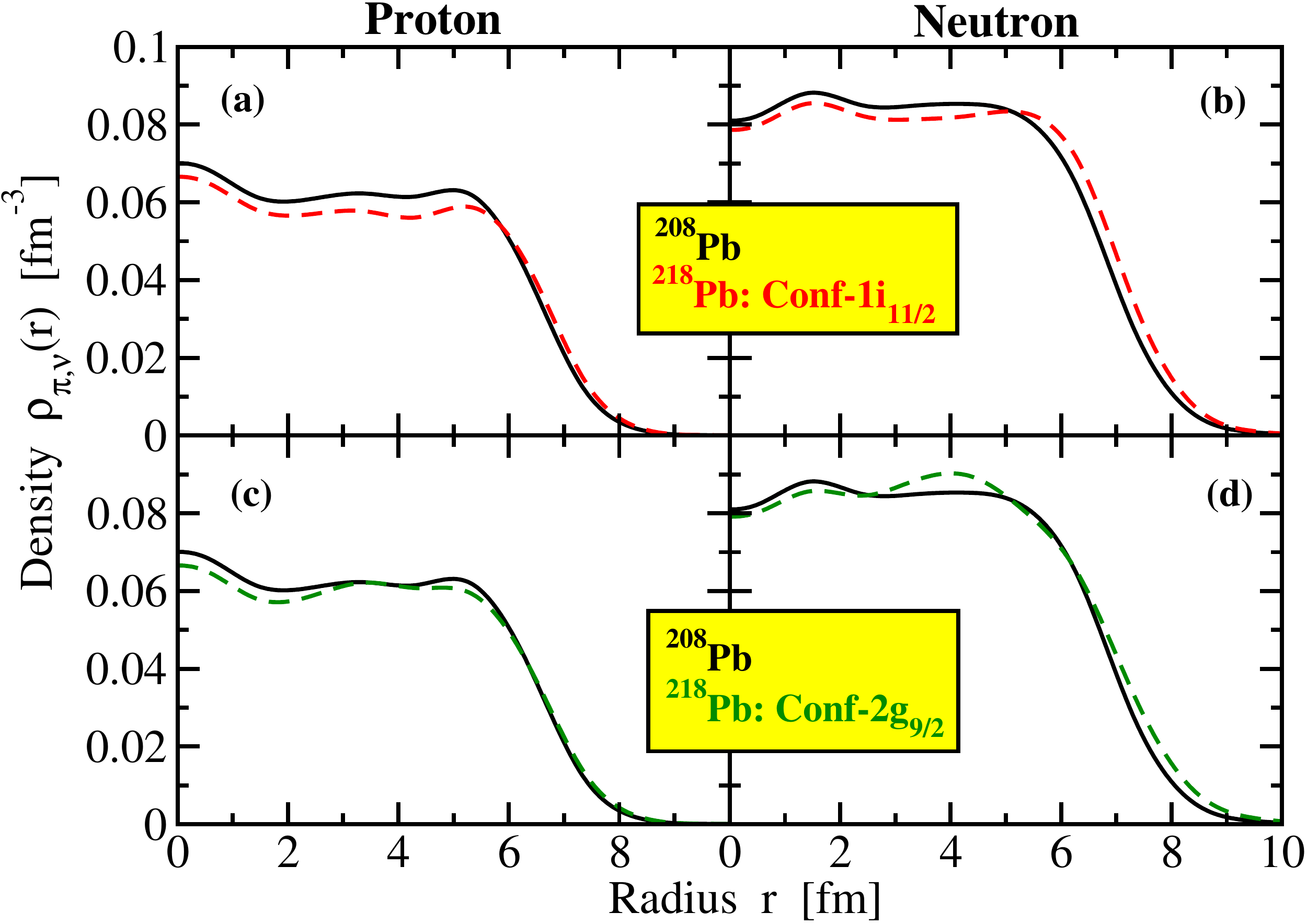}
\caption{Proton and neutron densities as a function of radial coordinate $r$
for indicated configurations.
\label{conf_density}
}
\end{figure}
%%%%%%%%%%%%%%%%%%%%%%%%%%%%%%%%%%%%%%%%%%%%%%%%

%%%%%%%%%%%%%%%%%%%%%%%%%%%%%%%%%%%%%%%%%%%%%%%
\begin{figure*}[ht]
\centering
\includegraphics[width=14.0cm]{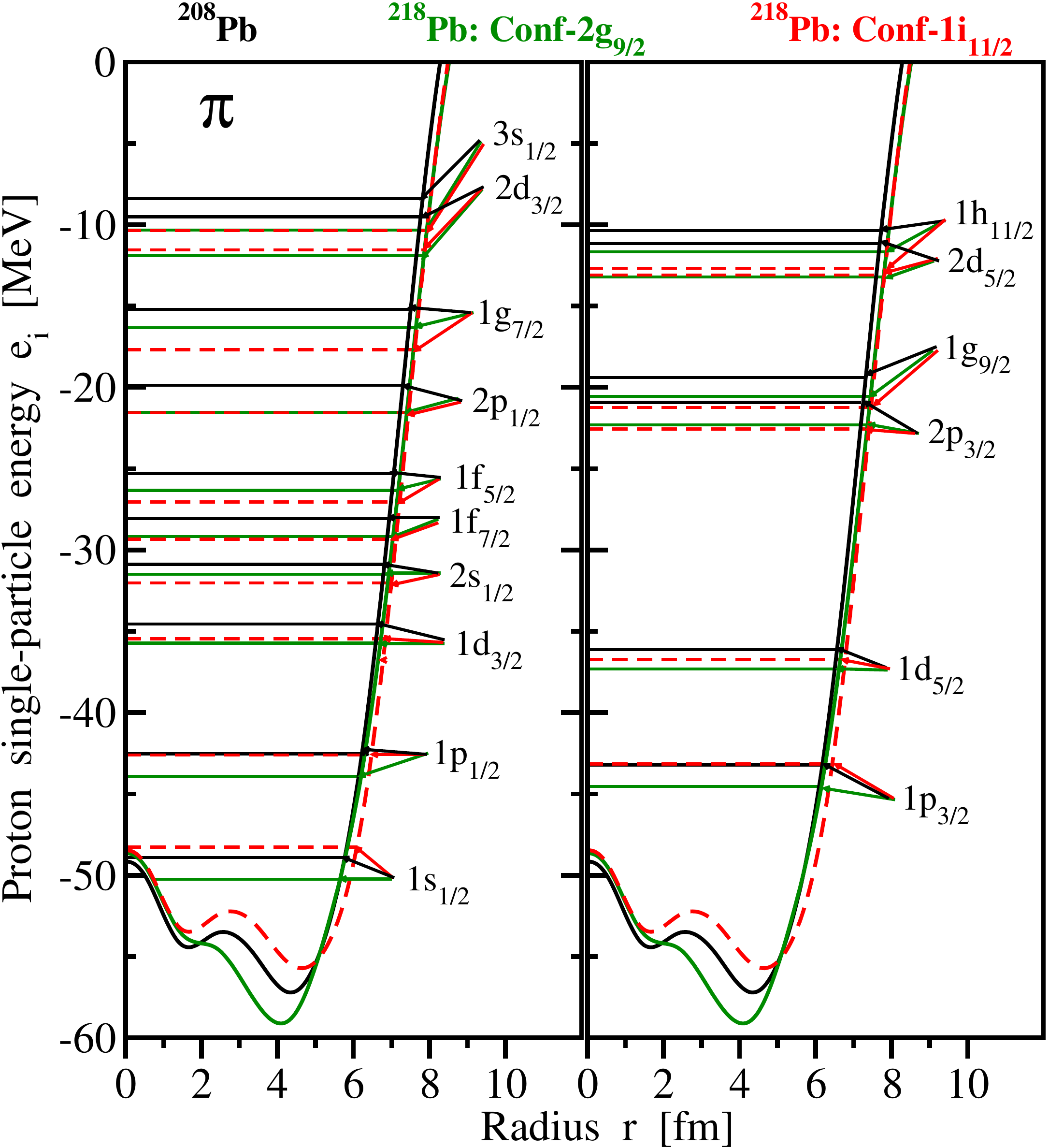}
\caption{The energies of proton single-particle states in the proton potentials 
of indicated nucleonic configurations. Note that only the states below the magic $Z=82$ 
shell closure are shown. The figure is split into two panels in order to make it more 
understandable  and  to avoid overlapping single-particle states.
\label{s-p-states}
}
\end{figure*}
%%%%%%%%%%%%%%%%%%%%%%%%%%%%%%%%%%%%%%%%%%%%%%%%

   Fig.\ \ref{conf_density} shows total proton and neutron densities of the configurations 
under study.  Neutron and proton densities of $^{208}$Pb serve as the references with 
respect of which the effect of the occupation of the neutron  $1i_{11/2}$ and  $2g_{9/2}$ subshells 
on total neutron and proton densities is discussed.  The occupation of the neutron $1i_{11/2}$ 
subshell in $^{218}$Pb ("Conf-1$i_{11/2}$") builds the neutron density mostly in the near-surface 
and  surface regions [see Fig.\ \ref{conf_density}(b)] since the single peak of its single-particle density is located at 
$r\approx 6.3$ fm (see Fig. 6(z) in Ref.\ \cite{PA.22}).  In contrast, the occupation of the neutron 
$2g_{9/2}$ subshell in $^{218}$Pb ("Conf-$2g_{9/2}$") builds the neutron density both in the 
subsurface and surface regions [see Fig.\ \ref{conf_density}(d)]. This is due to the nodal structure 
of the wave function of the $2g_{9/2}$ subshell which has two peaks in the single-particle density 
distribution: the large peak at $r\approx 4.4$ fm and the smaller one at  $r\approx 7.8$ fm (see 
Fig. 6(x) in Ref.\ \cite{PA.22}). 

    Because of the isovector force, which tries to keep the neutron and proton 
density profiles alike, the neutron density changes discussed above feed back into proton 
densities [see Figs.\ \ref{conf_density}(a) and (c)].  As compared with proton densities in 
$^{208}$Pb, the occupation of either neutron $1i_{11/2}$ or   $2g_{9/2}$ subshells leads 
to the reduction of the proton densities in the central and subsurface regions and increases 
the density in the surface region. These changes are more pronounced in the case of the 
occupation of  the neutron $1i_{11/2}$ subshell (the "Conf-1$i_{11/2}$" configuration in 
$^{218}$Pb).

%%%%%%%%%%%%%%%%%%%%%%%%%%%%%%%%%%%%%%%%%%%%%%%%
\begin{table*}[htb]
\caption{The contribution of different spherical subshells (column 7)  into the build-up 
of differential charge radii of  the "Conf-1$i_{11/2}$" configuration in $^{218}$Pb. The ground state 
configuration in $^{208}$Pb is used here as a reference. Proton subshells occupied below the $Z=82$
shell closure are shown in  column 1. Their  single-particle energies $e_i$ [in MeV] are shown in
columns 2 and 3 for two configurations under study.  Their proton single-particle rms radii 
$r_i^p$
are displayed in columns 4 and 5.  The column 6 
shows the change of proton single-particle radii 
$\delta r^p_i  = r^p_i (^{218}$Pb[Conf-$1i_{11/2}]) - r^p_i (^{208}$Pb).
Column 7 shows differential single-particle proton radii $\Delta \left<r^2\right>_i^{N,N'}$ 
of the single-particle state belonging to 
the $i$-th proton subshell. 
The overlaps of proton and neutron wave functions $<\Psi_n^k | \Psi_p^i>$
are shown in column 8. The results for spherical subshells with principal quantum number $n=1$ are shown in bold.
The total quantities given in the last line are calculated using Eq.\ (\ref{differ-sum}) (column 7) and 
equations  similar to it (columns 4, 5 and 6). }
\begin{tabular}{|c|c|c|c|c|c|c|c|}
\hline
 subshell           & $e_i$ ($^{208}$\rm{Pb}) & $e_i$ ($^{218}$\rm{Pb}) & $r^p_i$ ($^{208}$\rm{Pb}) & $r^p_i$ ($^{218}$\rm{Pb}) & $\delta r^p_i$  & 
      $\Delta \left < r^2 \right>_i^{N,N'}$ & $<\Psi_n^k | \Psi_p^i>$ \\ \hline
       1                &          2           &          3           &          4           &      5                 &         6       & 7    & 8  \\ \hline 
$1s_{1/2}$	& \bf{-48.905}	& \bf{-48.271}	& \bf{4.064254} & \bf{4.235252}  & \bf{0.170998}   &   \bf{1.419199}  &    \bf{0.570778}  \\
$1p_{3/2}$	& \bf{ -43.211}	& \bf{-43.156 }	& \bf{4.663197}	& \bf{4.846186}	  & \bf{0.182989}   &   \bf{1.740112}  &   \bf{0.729091} \\
$1p_{1/2}$	& \bf{-42.529}	& \bf{-42.598}   & \bf{	4.582763}& \bf{	4.771297}	  & \bf{0.188534}   &  \bf{1.763556}    & \bf{0.694871}  \\
$1d_{5/2}$	& \bf{-36.118}	& \bf{-36.727}	& \bf{5.105879}	& \bf{5.283008}	  & \bf{0.177129}   & \bf{ 1.840178}   & \bf{0.795503}   \\
$1d_{3/2}$	&\bf{ -34.559}	& \bf{-35.462}	& \bf{4.981504}	& \bf{ 5.159900} & \bf{0.178396}   &   \bf{1.809189}  & \bf{0.795142}     \\
$2s_{1/2}$	&  -30.886       &	-32.032	& 4.450498	&	4.528233	  &  0.077735        &   0.697962 & -0.638347     \\
$1f_{7/2}$ 	& \bf {-28.068}	&\bf{-29.330}	& \bf{5.479627}	& \bf{5.643146}	  & \bf{0.163519}  &  \bf{1.818783}   & \bf{0.899794}     \\
$1f_{5/2}$ 	& \bf{-25.298}	&\bf{-27.044}	& \bf{5.334254}	& \bf{5.490660}	  & \bf{0.156406}  & \bf{1.693085 }  & \bf{0.837178}   \\
$2p_{3/2}$	&	-20.924	&	-22.559	&  4.985584	&	 5.017296  &	0.031712     & 0.317213   & -0.571543 \\
$2p_{1/2}$	&	-19.865	&	 -21.566	&  5.004156	&	5.031929	  &	0.027773     & 0.278730   & -0.583835   \\
$1g_{9/2}$	& \bf{-19.396}	&\bf{-21.242 }   & \bf{5.816346}	& \bf{5.962968}  & \bf{0.146622}  & \bf{ 1.727100}  & \bf{ 0.910026}   \\
$1g_{7/2}$	& \bf{-15.205}   &\bf{-17.693}   & \bf{	5.682103}& \bf{	 5.807857} & \bf{0.125754}  & \bf{ 1.444908}  & \bf{0.932110}  \\
$2d_{5/2}$	&	-11.163	&	 -13.089 	&	5.522224   &	5.527597	  &	0.005373     & 0.059372  & -0.521033 \\
$1h_{11/2}$	& \bf{-10.360}   &\bf{-12.684  }	&\bf{	6.129798}	& \bf{6.257349}	  & \bf{0.127551}  &  \bf{1.579994} & \bf{ 0.975584}   \\
$2d_{3/2}$	&	-9.513	&	-11.537	&	5.580005	&  5.584444	  &	0.004439     & 0.049559  &  -0.554369  \\
$3s_{1/2}$	&	-8.405	&	-10.360	&5.489444	& 5.476911	  &	-0.012533    & -0.137438  & 0.448467  \\ \hline
Total &  & &5.450221&5.569299&0.119078& 1.312261 & \\\hline
\hline
\end{tabular}
\label{table-sp-radii_Pb_1i_11_2}
\end{table*}
%%%%%%%%%%%%%%%%%%%%%%%%%%%%%%%%%%%%%%%%%%%%%%%%

    Note that there are intricate and sometimes counterintuitive interplays between 
the changes in the densities and nucleonic potentials the origin of which was discussed in 
detail in Sec. V of Ref.\ \cite{PA.22}. For example, the density changes on going from the 
ground state configuration in $^{208}$Pb to the configurations in $^{218}$Pb show larger 
increase in the radial profile of neutron densities as compared with  proton ones
(see Fig.\ \ref{conf_density}).  In contrast, the respective changes in the radial 
profiles of proton potentials of the configurations in $^{218}$Pb with respect 
of that in $^{208}$Pb are larger than those for neutron potentials (see  
Fig.\ \ref{Pb_VPS_neu_pro}).

%%%%%%%%%%%%%%%%%%%%%%
\subsection{Proton single-particle states}
\label{self-sp-states}
%%%%%%%%%%%%%%%%%%%%%%

 Another consequence of the increase of neutron number on going from $^{208}$Pb to 
$^{218}$Pb is the lowering of the energies of the single-particle states in proton potential 
(see Fig.\ \ref{s-p-states}).  The energies of spherical subshells located at energies higher
than $-40$ MeV  in the "Conf-1$i_{11/2}$"  and "Conf-$2g_{9/2}$" configurations of 
$^{218}$Pb are lower than those in the ground state configuration of $^{208}$Pb. Such 
lowering implies some reduction of the rms proton radius as compared with the one at 
the energy of the subshell corresponding to the ground state configuration of $^{208}$Pb. 
However, as follows from further discussion the effect is rather marginal.

 Note  that  there is either small or no energy splitting between the energies of a given 
proton subshell calculated  in these two configurations of $^{218}$Pb  if the principal quantum
number $n$ of the subshell is either $n=2$ or $n=3$ (see Fig.\ \ref{s-p-states}). In contrast 
such energy splitting is typically  large for the subshells with $n=1$. These features are not very 
important for charge radii in the calculations without pairing. However, they are more
important in the calculations with pairing since the shifts of the energies of spherical 
subshells can affect their occupation probabilities in the vicinity of the Fermi level
(see discussion in Refs.\ \cite{Pb-Hg-radii-PRC.21,PAR.21}).

Deeply lying $1s_{1/2}$, $1p_{3/2}$ and $1p_{1/2}$ spherical subshells are affected 
by the properties and modifications (as compared with $^{208}$Pb one) of the bottom 
of the proton potential which shows the development of wine bottle potential features
(see Ref.\ \cite{PA.22}). These features are most/least pronounced in the 
"Conf-$2g_{9/2}$"/"Conf-1$i_{11/2}$" configurations of $^{218}$Pb (see Fig.\ 
\ref{s-p-states}).  As a consequence, for a given spherical subshell the calculated 
energies are almost the same in the ground state configuration of $^{208}$Pb
and in the "Conf-1$i_{11/2}$" configurations of $^{218}$Pb but significantly lower
for the "Conf-$2g_{9/2}$" configuration of $^{218}$Pb. Note that the presence
of classically forbidden region in proton potential at small radial coordinate $r$ 
can significantly  modify the density distribution of the $1s_{1/2}$ subshell in the 
"Conf-$2g_{9/2}$" configuration of $^{218}$Pb and can have some impact 
on density distribution of this subshell in other two configurations under study
(see detailed discussion in Sec. IV of Ref.\ \cite{PA.22}).

%%%%%%%%%%%%%%%%%%%%%%%%%%%%%%%%%%%%%%%%%%%%%%%%%%%
\begin{table*}[htb]
\caption{The same as in Table \ref{table-sp-radii_Pb_1i_11_2} but for the "Conf-$2g_{9/2}$" 
configuration in $^{218}$Pb. The results for spherical subshells with principal quantum 
number $n=2$ are shown in bold. }
\begin{tabular}{|c|c|c|c|c|c|c|c|}
\hline
 subshell  & $e_i$ ($^{208}$\rm{Pb}) & $e_i$ ($^{218}$\rm{Pb}) & $r^p_i$ ($^{208}$\rm{Pb}) & $r^p_i$ ($^{218}$\rm{Pb}) & $\delta r^p_i$ & $\Delta \left < r^2 \right>_i^{N,N'}$ & $<\Psi_n^k | \Psi_p^i>$ \\ \hline
       1                &          2           &          3           &          4             &      5                 &         6              & 7  & 8      \\ \hline 
$1s_{1/2}$	& -48.905         & -50.239	        & 4.064254          & 4.062288        & -0.001966        &  -0.015978    &0.399380\\
$1p_{3/2}$	&  -43.211	        & -44.539	        & 4.663197	  & 4.631773	    & -0.031424        &  -0.292086    &0.345824 \\
$1p_{1/2}$	& -42.529         & -43.915          & 4.582763         & 4.535377	    & -0.047386        & -0.432074     &0.373550 \\
$1d_{5/2}$	& -36.118         & -37.305          & 5.105879       	& 5.092180	    & -0.013699        &   -0.139704   &0.273309  \\
$1d_{3/2}$	&-34.559	        & -35.720	        & 4.981504          & 4.962466	    & -0.019038        & -0.189316     &0.282916   \\
$2s_{1/2}$	&  \bf{ -30.886} &\bf{-31.484}    & \bf{ 4.450498}   &\bf{4.565793}  & \bf{ 0.115295}  &   \bf{  1.039534}&\bf{0.182596}      \\
$1f_{7/2}$ 	& -28.068	        &-29.174	        & 5.479627          & 5.496088	    & 0.016461         & 0.180670     &0.119881 \\
$1f_{5/2}$ 	& -25.298	        & -26.329	        & 5.334254          & 5.363151	    & 0.028897         & 0.309126     &0.186386 \\
$2p_{3/2}$	& \bf{ -20.924}	& \bf{ -22.315 }	&  \bf{ 4.985584} &\bf{  5.158503} &	\bf{0.172919}  & \bf{ 1.754108 }    &\bf{0.475303}  \\
$2p_{1/2}$	& \bf{ -19.865}	& \bf{  -21.544}	& \bf{   5.004156}&\bf{ 5.170061}  &	\bf{0.165905}  & \bf{ 1.687954 }   &\bf{0.447216} \\
$1g_{9/2}$	& -19.396	        &-20.544           & 5.816346	  & 5.862341         & 0.045995         &  0.537159   &-0.000955\\
$1g_{7/2}$	& -15.205         &-16.330           & 5.682103         & 5.757601	    & 0.075498         & 0.863670    &-0.000587\\
$2d_{5/2}$	& \bf{ -11.163}  & \bf{ -13.214}	&\bf{ 5.522224}  &\bf{ 5.663334}   &	\bf{0.14111}    & \bf{ 1.578392 }   &\bf{0.754258}  \\
$1h_{11/2}$	& -10.360         &-11.659           &6.129798	 & 6.199679	    & 0.069881         &   0.861599   &-0.182783\\
$2d_{3/2}$	& \bf{ -9.513}    & \bf{ -11.887}	&\bf{ 5.580005} & \bf{ 5.686162	}   &	\bf{0.106157}  &  \bf{1.195982  } &\bf{0.719801}  \\
$3s_{1/2}$	&	-8.405      &	-10.335	&5.489444	 & 5.540033	    &	0.050589        & 0.557970    &-0.7388235  \\ \hline
Total &  & &5.450221&5.505978&0.055757& 0.6110 &\\
\hline
\hline
\end{tabular}
\label{table-sp-radii_Pb_2g9_2}
\end{table*}
%%%%%%%%%%%%%%%%%%%%%%%%%%%%%%%%%%%%%%%%%%%%%

%%%%%%%%%%%%%%%%%%%%%%%%%%%%%%%%%%%%%%%%%%%%%%%%
\begin{figure*}[htb]
\centering
\includegraphics[width=15.4cm]{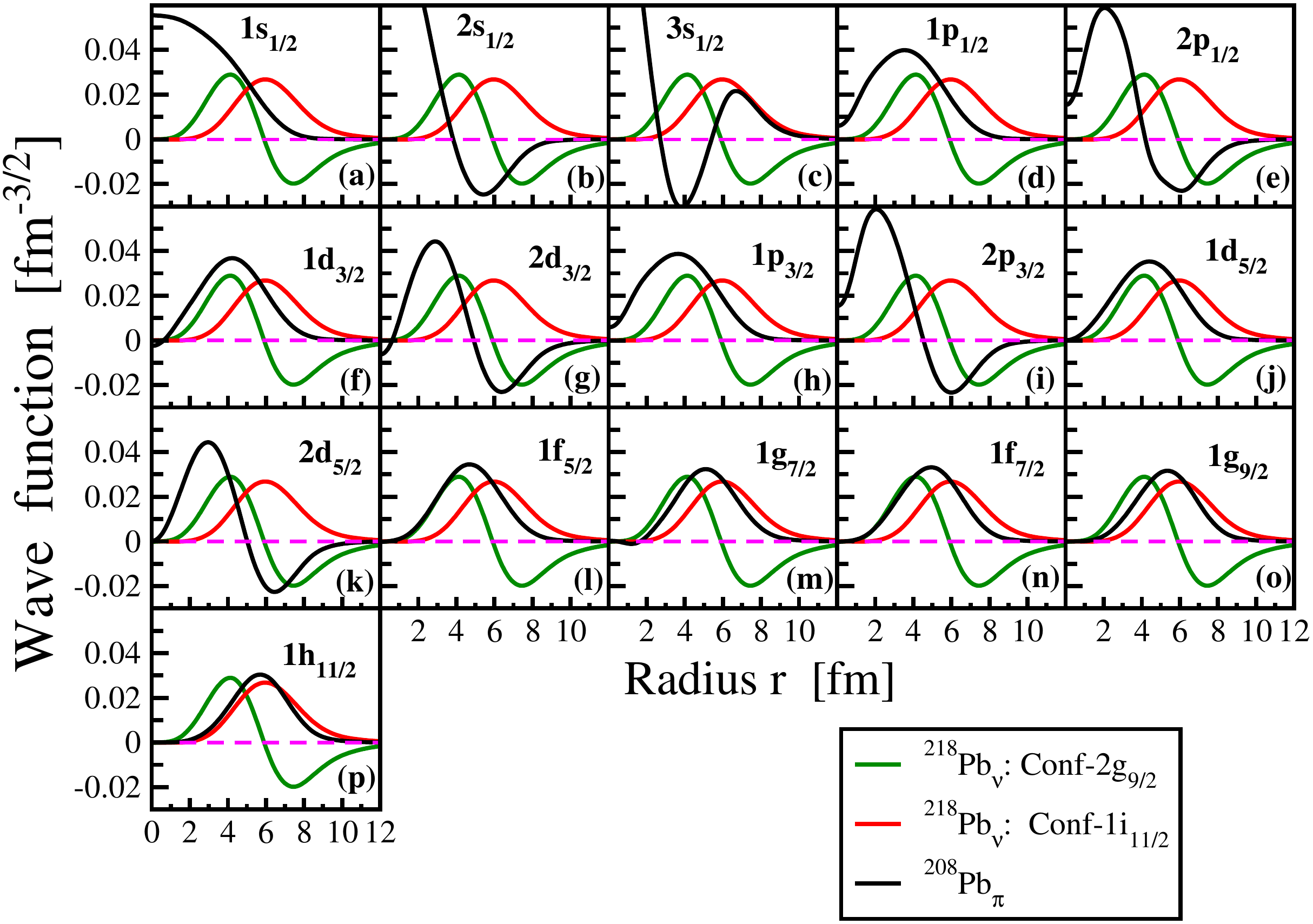}
\caption{Single-particle wave functions of proton (black curves) spherical subshells 
in $^{208}$Pb nucleus compared with those of the neutron $2g_{9/2}$ (green lines) and 
$1i_{11/2}$ (red lines) subshells. Note that the latter two states are shown in each 
panel. Proton subshell labels are shown on each panel. 
\label{Pb_wavefunctions}
}
\end{figure*}
%%%%%%%%%%%%%%%%%%%%%%%%%%%%%%%%%%%%%%%%%%%%%%%% 

%%%%%%%%%%%%%%%%%%%%%%%%%%%%%%%%
\subsection{Differential charge radii: microscopic origin of the kinks}
%%%%%%%%%%%%%%%%%%%%%%%%%%%%%%%%  

   It is well established in different model calculations that dominant or
significant occupation of the neutron $1i_{11/2}$ subshell above the
$N=126$ shell closure is critical for explaining the kink in differential 
charge radii at neutron $N=126$  shell closure 
\cite{SLR.93,RF.95,GSR.13,NI.15,Pb-Hg-charge-radii-PRL.21,Pb-Hg-radii-PRC.21,PAR.21}. 
The functionals (typically CEDFs) in which the $\nu 1i_{11/2}$ subshell is 
located below  the $\nu 2g_{9/2}$ subshell reasonably well describe this 
kink and the slopes of differential charge radii below and above $N=126$. 
In contrast, the functionals (typically non-relativistic  ones) with opposite
location of two subshells either fail to describe the kink or significantly underestimate
its magnitude.

  This difference in the slope of differential charge radii above the $N=126$ 
shell closure as emerging due to the occupation of above mentioned neutron spherical 
subshells has been discussed earlier within unpaired RMF calculations with 
the NL3* CEDF (see discussion of Fig.\ 4 in Ref.\ \cite{PAR.21}). The 
differential charge radii $\delta \left < r^2 \right>_p^{136,126}$ are 
1.31  fm$^2$ and 0.61 fm$^2$ for the "Conf-1$i_{11/2}$" and 
"Conf-$2g_{9/2}$" configurations  of $^{218}$Pb, respectively.
However, the microscopic  origin of this significant difference has not been investigated.  
To fill the gap in our knowledge, we show in Tables \ref{table-sp-radii_Pb_1i_11_2} and 
\ref{table-sp-radii_Pb_2g9_2}  the contributions of different spherical subshells into 
build-up of differential charge radii of the "Conf-1$i_{11/2}$"  and "Conf-$2g_{9/2}$" 
configurations of $^{218}$Pb. Note that  for the sake of comparison with single-particle 
wave functions and densities we also consider proton single-particle rms radii $r^p_i$
in these tables.

  Table \ref{table-sp-radii_Pb_1i_11_2}  clearly shows that when the neutron
$1i_{11/2}$ subshell is occupied in $^{218}$Pb  the  largest changes in proton 
single-particle rms radii  $\delta r^p_i$ take place for the proton subshells 
with principal quantum number $n=1$. The $\delta r^p_i$ values for the $s$, $p$ 
and $d$ $n=2$ subshells are smaller than the average $\delta r^p_i$ value over 
the $n=1$ subshells  by a factor of approximately 2, 5 and 25, respectively. Proton single-particle 
rms radius of the $3s_{1/2}$ subshell even decreases on transition from the ground 
state configuration of $^{208}$Pb to the "Conf-1$i_{11/2}$" configuration of 
$^{218}$Pb.  All these changes are reflected in differential single-particle proton 
radii $\Delta \left<r^2\right>_i^{N,N'}$ (see column 7 of Table \ref{table-sp-radii_Pb_1i_11_2})
so that  96.7\% of differential charge radius $\delta \left<r^2\right>_p^{136,126}= 1.31$ fm$^2$ 
[see Eq.\ (\ref{diff-radii})]  of the "Conf-1$i_{11/2}$" configuration in $^{218}$Pb with respect 
of the ground state configuration in $^{208}$Pb  are built by the $n=1$ proton subshells.

 The situation drastically changes when the neutron $2g_{9/2}$ subshell is 
occupied in $^{218}$Pb [configuration "Conf-$2g_{9/2}$"] (see Table 
\ref{table-sp-radii_Pb_2g9_2}).  In this case, the largest $\delta r^p_i$
values are seen for the $2s_{1/2}$, $2p_{3/2}$, $2p_{1/2}$, $2d_{5/2}$
and $2d_{3/2}$ proton subshells.  The proton rms radii of  low lying  $1s_{1/2}$, 
$1p_{3/2}$, $1p_{1/2}$, $1d_{5/2}$ and $1d_{3/2}$ subshells even decrease on transition 
from  the ground state configuration of $^{208}$Pb to the "Conf-2$g_{9/2}$"  
configuration of $^{218}$Pb. This is the consequence of self-consistency effects 
discussed in Secs.\ \ref{self-pot-den} and \ref{self-sp-states} and the effects 
discussed in Sec.\ \ref{mic-pull}. In addition, the increase of  proton rms radii 
is rather modest for remaining $n=1$ subshells and for the $3s_{1/2}$ subshell.
All these changes are reflected in differential single-particle proton 
radii $\Delta \left<r^2\right>_i^{N,N'}$ (see column 7 of Table \ref{table-sp-radii_Pb_2g9_2})
so that  53.3\% of differential charge radius $\delta \left<r^2\right>_p^{136,126}= 0.61 $ fm$^2$ 
of two configurations under study are built by the $n=2$ proton subshells. This is despite the 
low multiplicity of the occupied $n=2$ subshells which represent only approximately 
22\% of the occupied single-particle states of the $Z=82$ core.

%%%%%%%%%%%%%%%%%%%%%%%%%%%%%%%%%%%%%%%%%%%%%%%%
\begin{figure}[htb]
\centering
\includegraphics[width=8.5cm]{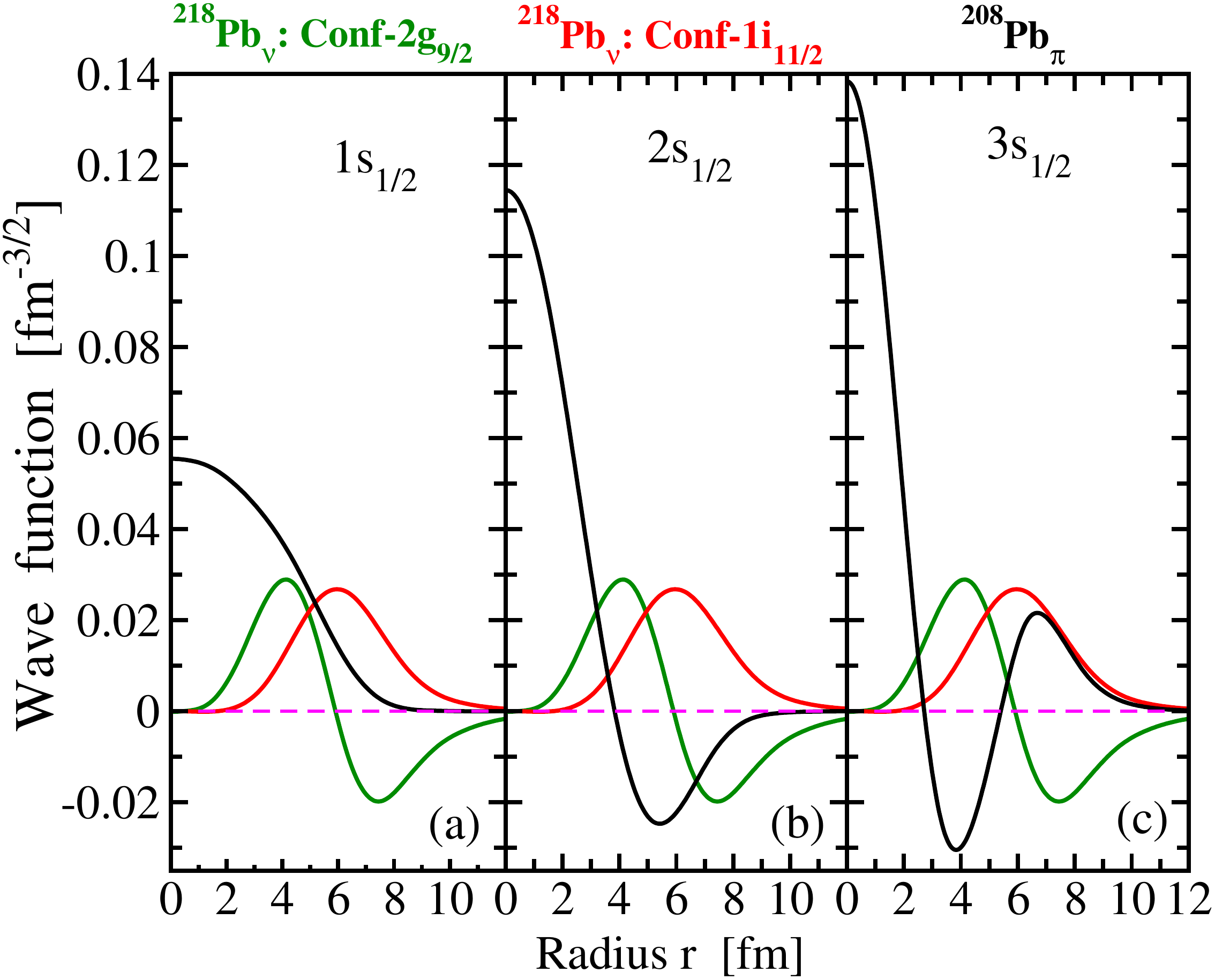}
\caption{The same as Fig.\ \ref{Pb_wavefunctions} but only for the  proton $s$ states
and with enlarged range on vertical axis.
\label{Pb_wavefunctions_s_states}
}
\end{figure}
%%%%%%%%%%%%%%%%%%%%%%%%%%%%%%%%%%%%%%%%%%%%%%%% 

The impact of the occupation of neutron $2g_{9/2}$ subshell on proton single-particle rms radii 
of the $n=2$ proton subshells (see Table \ref{table-sp-radii_Pb_2g9_2}) is on average comparable 
to the one of neutron $1i_{11/2}$ subshell on proton single-particle rms radii of the $n=1$ proton 
subshells (see Table \ref{table-sp-radii_Pb_1i_11_2}).    Thus, other factors have to 
be involved to explain large difference in differential charge radii of the "Conf-1$i_{11/2}$" and
"Conf-2$g_{9/2}$" configurations in $^{218}$Pb. Indeed, the analysis of Eq.\ (\ref{differ-sum}) 
clearly indicates that differential charge radii between two isotopes are defined not only by 
the differential single-particle radii $\Delta \left<r^2\right>_i^{N,N'}$ of occupied single-particle
states 

but  also by the abundance of the subshells with a given $n$ among occupied
subshells and their multiplicity $m_i$.  Low $n$ subshells are most abundant in any
nucleonic potential (see Refs.\ \cite{NilRag-book,PA.22}). Indeed,  there are
10 $n=1$, 5 $n=2$ and 1 $n=3$ occupied subshells in the $Z=82$ core of the
Pb isotopes (see Fig.\ \ref{s-p-states}). In addition the $n=1$ subshells have 
the highest multiplicity $m_{max}$ among the occupied subshells: $m_{max}=12$, 
6 and 2 for the $n=1$, 2 and 3 subshells, respectively. As a result, 62, 18 and 2 
protons of the $Z=82$ core are located in the $n=1$, 2 and 3 subshells, 
respectively. The combination of all above mentioned in this subsection factors allows to explain  large difference 
in differential charge radii of the "Conf-1$i_{11/2}$" and "Conf-2$g_{9/2}$" 
configurations in $^{218}$Pb.

%%%%%%%%%%%%%%%%%%%%%%%%%%%%%%%%%%%%%%%%
\subsection{Microscopic origin of the pull of neutron subshells on proton ones}
\label{mic-pull}
%%%%%%%%%%%%%%%%%%%%%%%%%%%%%%%%%%%%%%%%

 In order to better understand the state dependence of the pull provided by
a neutron in a given state on the proton in the $nlj$ subshell, Figs.\ 
\ref{Pb_wavefunctions} and \ref{Pb_wavefunctions_s_states} compare the 
proton wave functions of all occupied proton subshells in the $Z=82$ core of 
the $^{208}$Pb nucleus with the neutron wave functions  of the neutron 
$1i_{11/2}$ and $2g_{9/2}$  subshells  calculated in the  "Conf-1$i_{11/2}$" 
and "Conf-2$g_{9/2}$"  configurations of $^{218}$Pb, respectively.
In addition, the overlap of respective proton and neutron 
wave functions defined as
\begin{eqnarray}
<\Psi_n^k | \Psi_p^i> = \int \Psi_n^k ({\vec r}\,) \Psi_p^i ({\vec r}\,)  d^3r
\end{eqnarray}
is presented in the last columns of Tables \ref{table-sp-radii_Pb_1i_11_2}
and  \ref{table-sp-radii_Pb_2g9_2}.  Here proton state index $i$ runs over 
all occupied proton subshells while neutron index $k$ is equal either to 
$k=1i_{11/2}$ or $k=2g_{9/2}$.  Positive (negative) values of these overlaps
indicate that the wave functions $\Psi_n^k$ and $\Psi_p^i$ are spatially 
mostly in phase (out of phase). 

  Let us first consider the overlaps of proton wave functions with the neutron 
$\nu 1i_{11/2}$ one [see Table \ref{table-sp-radii_Pb_1i_11_2}]. The largest 
overlap exists for the $\pi 1h_{11/2}$ state ($<\Psi_n^k | \Psi_p^i>=0.98$). Indeed,
these two states  have the wave functions which are most similar among
considered cases [see Fig.\ \ref{Pb_wavefunctions}(p) and compare it with other
panels of this figure].  The degree of the similarity ($<\Psi_n^k | \Psi_p^i> \approx 0.92$)
of the wave function of the neutron $\nu 1i_{11/2}$ subshell is  somewhat smaller  
with the wave functions of  the proton $\pi 1g_{7/2}$ and $\pi 1g_{9/2}$ subshells 
[see Fig.\ \ref{Pb_wavefunctions}(m) and (o)]. With decreasing the single-particle
energy of spherical $n=1$ proton subshell the degree of the similarity between 
neutron and proton wave functions given by $<\Psi_n^k | \Psi_p^i>$ decreases
but still remains high (see the last column of Table \ref{table-sp-radii_Pb_1i_11_2}
and  Figs.\ \ref{Pb_wavefunctions}(l),  (j), (h), (f), (d) and (a)]. 
Among the proton $n=1$ subshells the lowest overlap $<\Psi_n^k | \Psi_p^i>=0.57$
exists for proton $1s_{1/2}$ subshell which is the only $n=1$ subshell with the
maximum of the wave function at the center of nucleus
[see Fig.\ \ref{Pb_wavefunctions}(a)].

   The situation is completely different for the proton subshells with $n=2$: the 
evolution of their wave functions as a function of radial coordinate $r$ is mostly 
out of phase with that of the wave function of the proton $1i_{11/2}$ subshell [see 
Figs.\ \ref{Pb_wavefunctions}(b), (e), (g), (i) and (k)]. This is due to the differences 
in the nodal structure of these wave functions. As a consequence, large negative 
overlaps $<\Psi_n^k | \Psi_p^i>$ exist for these pairs of the  states (see last column of Table 
\ref{table-sp-radii_Pb_1i_11_2}). There are large differences due to underlying 
nodal structure between the wave functions of the proton $3s_{1/2}$ and neutron 
$1i_{11/2}$ subshells (see Fig.\ \ref{Pb_wavefunctions}(c)). However, the overlap 
$<\Psi_n^k | \Psi_p^i>$ for this pair of the subshells is positive due to the fact that 
the overlap is dominated by the behavior of the wave functions at large radial 
coordinates.

 %%%%%%%%%%%%%%%%%%%%%%%%%%%%%%%%%%%%%%%%%%%%%%%%
\begin{figure}[htb]
\centering
\includegraphics[width=8.5cm]{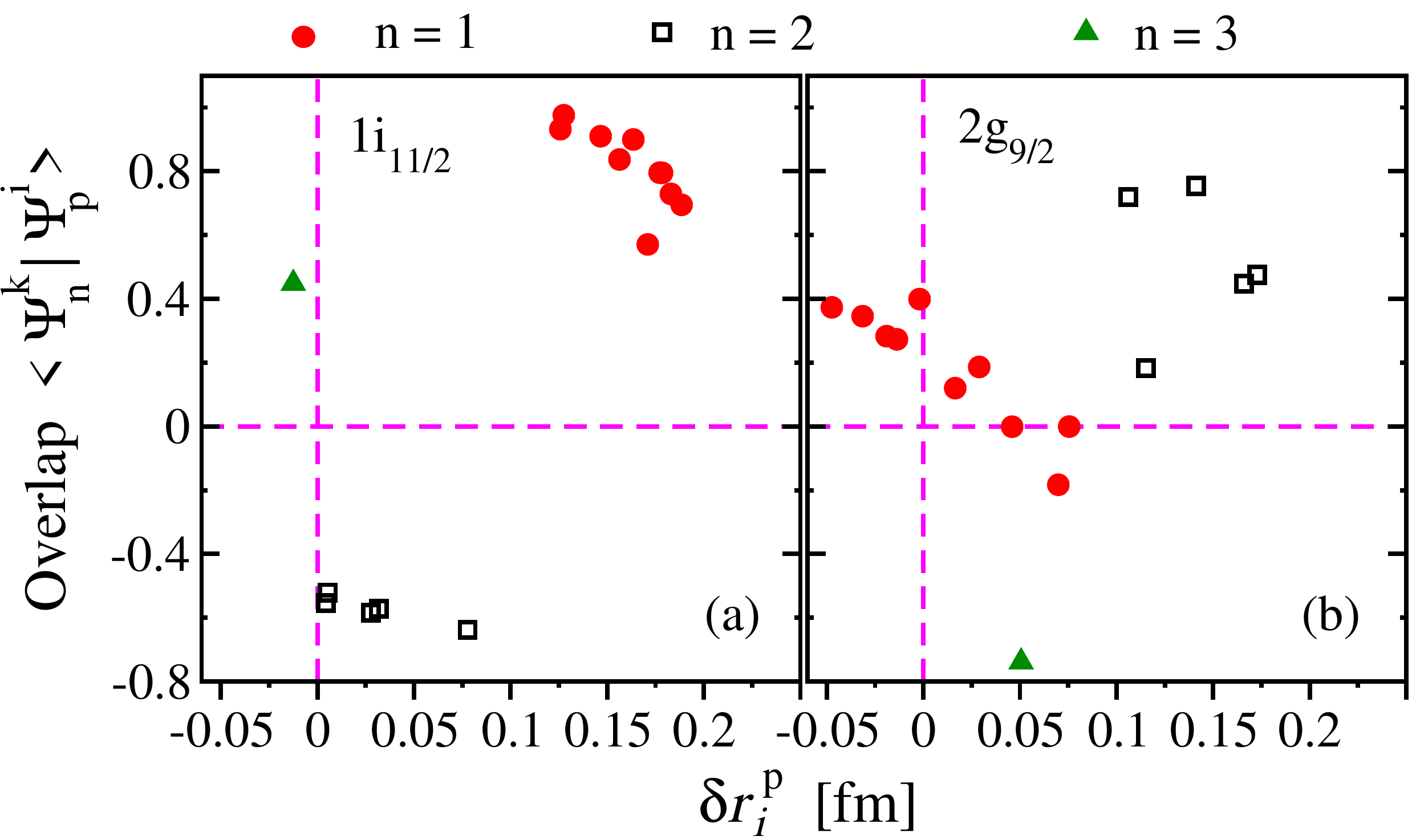}
\caption{The correlations between the overlaps $<\Psi_n^k | \Psi_p^i>$ and the changes 
of proton single-particle radii $\delta r^p_i$ for proton subshells of the $Z=82$ core. Solid red
circles, open black squares and green triangles are used for the $n=1$, $n=2$ and $n=3$
proton subshells. Panels (a) and (b) show the results when neutron $1i_{11/2}$ and
$2g_{9/2}$ subshells are occupied in the configurations of the $^{218}$Pb nucleus,
respectively.
\label{over-vs-radii-change}
}
\end{figure}
%%%%%%%%%%%%%%%%%%%%%%%%%%%%%%%%%%%%%%%%%%%%%%%%

 %%%%%%%%%%%%%%%%%%%%%%%%%%%%%%%%%%%%%%%%%%%%%%%%
\begin{figure}[htb]
\centering
\includegraphics[width=8.5cm]{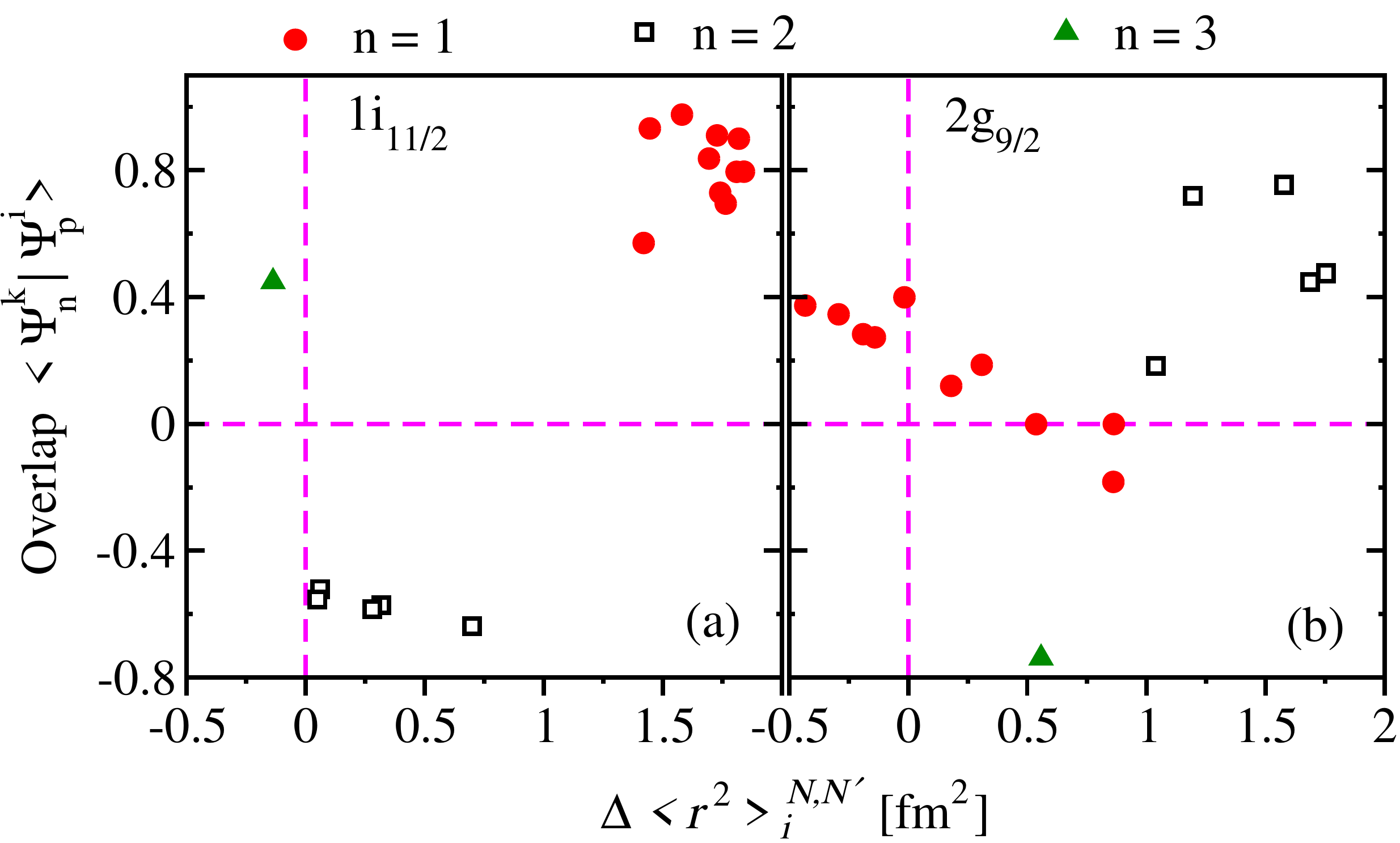}
\caption{The same as in Fig.\ \ref{over-vs-radii-change} but for the correlations 
between the overlaps $<\Psi_n^k | \Psi_p^i>$ and differential single-particle proton 
radii $\Delta \left< r^2 \right>_i^{N,N'}$.
\label{over-vs-diff-radii}
}
\end{figure}
%%%%%%%%%%%%%%%%%%%%%%%%%%%%%%%%%%%%%%%%%%%%%%%%

  The occupation of the neutron $2g_{9/2}$ subshell leads to completely different
pattern of behavior (see Figs.\ \ref{Pb_wavefunctions} and \ref{Pb_wavefunctions_s_states} and Table 
\ref{table-sp-radii_Pb_2g9_2}).  The largest overlaps exist for the $n=2$ 
proton subshells: the only exception is the overlap which includes proton $2s_{1/2}$ subshell
which has a maximum of its wave function at $r=0$.  These overlaps become smaller
or even negative for the cases which include $n=1$ and $n=3$ proton subshells
(see the last column in Table  \ref{table-sp-radii_Pb_2g9_2}).

%%%%%%%%%%%%%%%%%%%%%%%%%%%%%%%%%%%%%%%%%%%%%%%%
\begin{figure*}[htb]
\centering
\includegraphics[width=15.4cm]{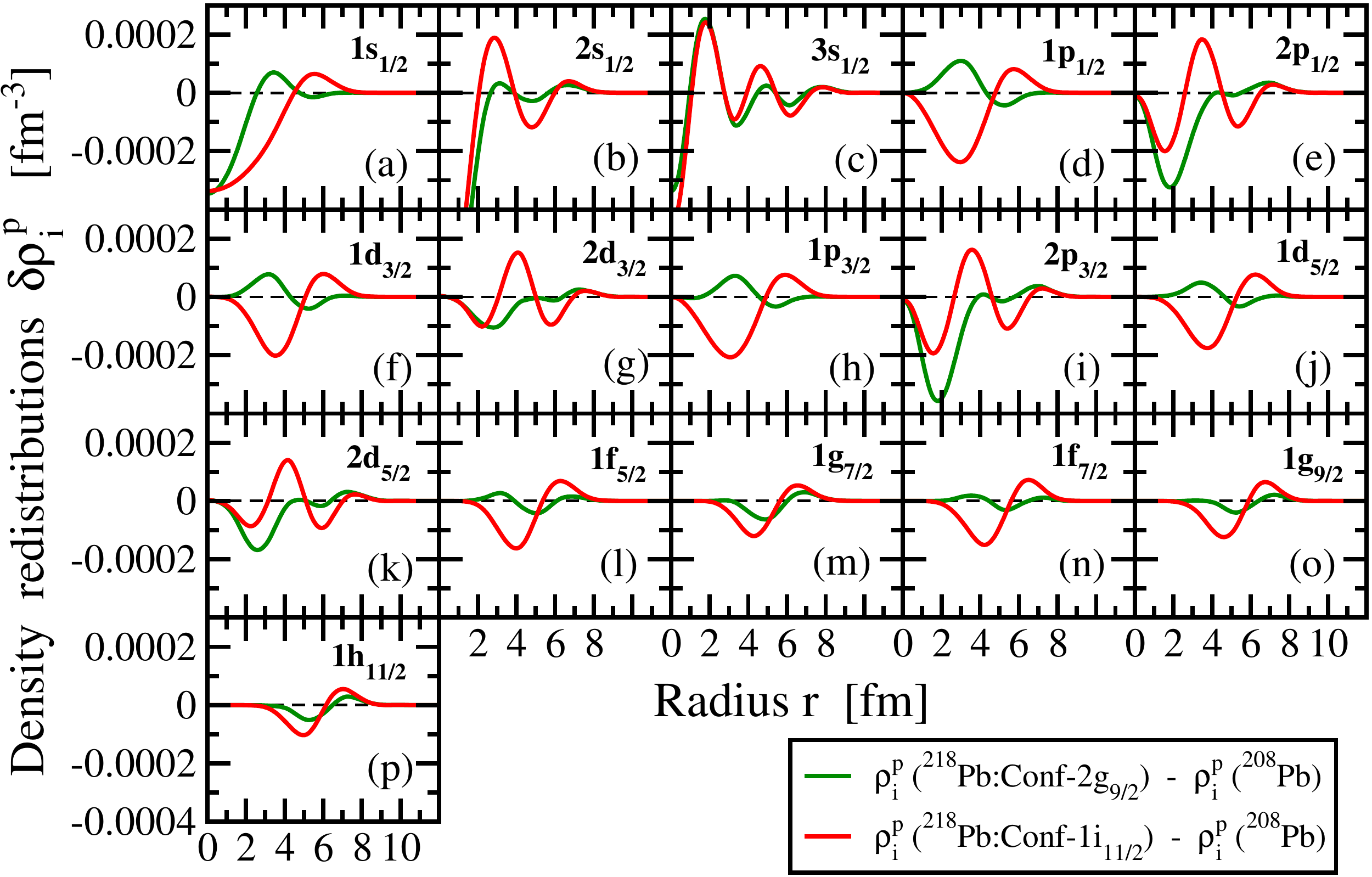}
\caption{Proton  single-particle  density redistributions $\delta \rho_i^p(r)$ caused by the 
occupation of indicated neutron subshells in $^{218}$Pb (see text for details).
\label{Pb_SPD_Difference_Pro}
}
\end{figure*}
%%%%%%%%%%%%%%%%%%%%%%%%%%%%%%%%%%%%%%%%%%%%%%%%%%%

  Fig.\ \ref{over-vs-radii-change} shows the correlations between the overlaps
$<\Psi_n^k | \Psi_p^i>$ for the neutrons in the $1i_{11/2}$ and $2g_{9/2}$ subshells 
and the proton subshells occupied in the $Z=82$ core and the changes in 
single-particle proton rms radii $\delta r_i^p$ of these subshells triggered by 
the occupation of respective neutron subshells.  In general, the largest 
$\delta r_i^p$ values appear for the proton subshells which have the same
principal quantum number $n$ as occupied neutron subshell. This also
corresponds to the largest positive overlaps $<\Psi_n^k | \Psi_p^i>$. Small
or negative overlaps, which correspond to the case of different principal
quantum numbers $n$ of proton and neutron subshells, typically lead to 
relatively small  $\delta r_i^p$ values.
 
     These correlations are very pronounced in the case of the occupation 
of the neutron $1i_{11/2}$ subshell since its wave function has a simple structure with 
a single maximum at $r\approx 6$ fm (see Fig.\ \ref{Pb_wavefunctions}). Significant 
changes in the single-particle rms radius are seen for proton $n=1$  subshells 
which have large overlaps $<\Psi_n^k | \Psi_p^i>$ but rather small  $\delta r_i^p$ 
values exist for the proton $n=2$ subshells which have negative overlaps (see Fig.\ 
\ref{over-vs-radii-change} and Table \ref{table-sp-radii_Pb_1i_11_2}).

   Such correlations are somewhat less pronounced in the case of the occupation of the 
$2g_{9/2}$ neutron subshell the wave function of which has maximum at $r\approx 4$ 
fm and minimum at $r=7.5$ fm (see Fig.\ \ref{Pb_wavefunctions}). The largest overlaps 
$<\Psi_n^k | \Psi_p^i>$ are seen with proton $n=2$ $2p_{3/2}$, $2p_{1/2}$, $2d_{5/2}$  
and $2d_{3/2}$ subshells which produce the largest changes in the proton single-particle 
rms radii (see Fig.\ \ref{over-vs-radii-change} and Table  \ref{table-sp-radii_Pb_2g9_2}). 
Smaller and sometimes negative changes in the proton single-particle rms radii are produced
for the $n=1$ and $n=3$ proton subshells when neutron $2g_{9/2}$ subshell becomes 
occupied.

 Note that similar to above discussed correlations are also seen between $<\Psi_n^k | \Psi_p^i>$
and differential single-particle proton radii $\Delta \left< r^2 \right>_i^{N,N'}$ (see Fig.\ 
\ref{over-vs-diff-radii}) with the latter quantities defining the differential charge radius between
two isotopes [see Eq. (\ref{differ-sum})].

   The absolute values of the  $<\Psi_n^k | \Psi_p^i>$ overlaps presented in 
column 8 of Tables \ref{table-sp-radii_Pb_1i_11_2} and \ref{table-sp-radii_Pb_2g9_2} 
are similar to those obtained in Skyrme DFT calculations with NRAPRii energy density 
functional (see Fig.\ 5 of Ref.\ \cite{GSR.13}).  This clearly indicates a similar mechanism 
of the buildup of differential charge radii in non-relativistic and covariant DFTs.
     
    Because of the dependence of charge radii and their changes on proton
single-particle densities (see Sec.\ \ref{Theory}), a deeper microscopic insight is provided 
by the analysis of the redistributions of the proton  single-particle  densities defined as
\begin{eqnarray}
\delta \rho_i^p(r) = \rho_i^p(r) [^{218}{\rm Pb}-{\rm conf}] - \rho_i^p(r) [^{208}{\rm Pb}] 
\end{eqnarray}
when different neutron subshells are occupied in $^{218}$Pb. 
Since the single-particle density is normalized to unity [see Eq.\ (\ref{norm-cond})], the 
addition of neutron(s) to the $^{208}$Pb nucleus will  only lead to redistribution of the 
proton single-particle density under the condition that
\begin{eqnarray}
\int r^2 \delta \rho_i^p(r) = 0
\end{eqnarray}

   These redistributions are clearly seen in Fig.\ \ref{Pb_SPD_Difference_Pro}. Let us 
consider as an example the changes in the single-particle proton densities of the 
$1p_{3/2}$ subshell [see Fig.\ \ref{Pb_SPD_Difference_Pro}(h)].  The occupation of 
the neutrons in the $2g_{9/2}$ subshell (the "Conf-2$g_{9/2}$" configuration in 
$^{218}$Pb) leads to an increase of proton densities of the $1p_{3/2}$ subshell at
$r\approx 3.2$ fm and their decrease at  $r\approx 5.5$ fm as compared with that
in the ground state configuration of $^{208}$Pb [green line in Fig.\ 
\ref{Pb_SPD_Difference_Pro}(h)]. In contrast, the occupation of the 
neutrons in the $1i_{11/2}$ subshell (the "Conf-1$i_{11/2}$" configuration in  $^{218}$Pb)
has an opposite effect: it leads to the decrease of proton densities of the $1p_{3/2}$ 
subshell at $r\approx 3$ fm and their increase\footnote{The reader should not be 
confused by larger density changes at low radial coordinates as compared with 
those at larger values of $r$. This is because the density plots as a function of radial 
coordinate tend to overemphasize the importance of the central region since they 
ignore the  fact that the number of particles $dn$ in a spherical shell of thickness 
$dr$ is given  by $4\pi r^2 \rho(r)dr$ (see example in Sec.\ III of Ref.\ \cite{PA.22}).}
at  $r\approx 6$ fm [red line in Fig.\ \ref{Pb_SPD_Difference_Pro}(h)]. Note that the 
number of the oscillations of density redistributions increases  with the increase of principal 
quantum number $n$ of the proton and neutron subshells involved (compare, for example, 
the panels (c) and (h) of Fig.\ \ref{Pb_SPD_Difference_Pro}).

    For almost all proton subshells of the $Z=82$ core the occupation of the $1i_{11/2}$ 
or $2g_{9/2}$ neutron subshells leads to drastically different redistributions of proton 
single-particle densities which are frequently out of phase of each others as a function of radial 
coordinate $r$ (see Fig.\ \ref{Pb_SPD_Difference_Pro}). Thus, for a given proton subshell  
this leads to different changes in the proton single-particle radii $\delta r_i^p$ and substantial 
differences in differential  single-particle radii $\Delta \left< r^2 \right>_i^{N,N'}$ (compare Tables 
\ref{table-sp-radii_Pb_1i_11_2} and  \ref{table-sp-radii_Pb_2g9_2}).

%%%%%%%%%%%%%%%%%%%%%%%%%%%%%%%%
\section{General observations}
\label{gen-observation}
%%%%%%%%%%%%%%%%%%%%%%%%%%%%%%%%  

%%%%%%%%%%%%%%%%%%%%%%%%%%%  
\subsection{The origin of differential charge radii}
\label{gen-origin}
%%%%%%%%%%%%%%%%%%%%%%%%%%%

   The detailed analysis presented in Sec.\ \ref{self-const} clearly indicates that 
the change of differential charge radius with increasing neutron number has
a single-particle origin. Added neutron interacts with proton in a given
subshell and this
leads to a redistribution of single-particle densities of occupied proton
states which in turn modifies the charge radii. The outcome of this 
interaction depends on the relative properties of the wave functions of
interacting proton and neutron. Large positive overlap $<\Psi_n^k | \Psi_p^i>$ 
between their wave functions leads to a substantial increase of differential 
single-particle charge radius of proton subshell in which interacting proton 
is located.  In contrast, small or negative overlap of their wave functions 
typically leads to either small increase or even decrease in   differential 
single-particle charge radius.

    Ref.\ \cite{GSR.13} has discussed the pull on the proton states provided 
by the occupation of neutron states and resulting change in charge radii
as emerging from strong nuclear symmetry energy. However, our detailed 
analysis clearly indicates that the changes in charge radii with increasing 
neutron number are governed by the proton-neutron interaction. 
     
  The buildup of  differential charge radii $\delta \left< r^2 \right>_p^{N,N'}$
between two isotopes is also a collective phenomenon since all occupied proton 
states contribute to it [see Eq.\ (\ref{differ-sum})] and none of these states
provides a dominant contribution to $\delta \left< r^2 \right>_p^{N,N'}$ (see 
Tables \ref{table-sp-radii_Pb_1i_11_2} and \ref{table-sp-radii_Pb_2g9_2}).

%%%%%%%%%%%%%%%%%%%%%%%%%%%%%%%%%%%%  
\subsection{Applicability of alternative models to the description of
                    differential charge radii in deformed nuclei}
%%%%%%%%%%%%%%%%%%%%%%%%%%%%%%%%%%%%
  
   As reviewed in the introduction of Ref.\ \cite{PAR.21},  the absolute majority of
the studies of differential  charge radii has been performed either in DFT or in ab initio
approaches.  However, it is well known that DFT models have some deficiencies
in the description of spectroscopic properties related to the energies  of the single-particle states
and their wave functions  \cite{BQM.07,CSB.10,AS.11,DABRS.15,AL.15}. Moreover,
the performance of ab initio models in the description of single-particle spectra in odd-$A$
nuclei is comparable with that for the DFT models but such calculations are available 
only for light nuclei (see Refs.\  \cite{HHJMP.12,JSSHN.16,JEHNS.14,BHHSBCLR.14}).
 In contrast, spherical
shell models with empirical interactions provide a better description of experimental spectroscopic
data in spherical nuclei  located in the vicinity  of doubly magic nuclei and  microscopic+macroscopic
(mic+mac) models based on phenomenological potentials such as the Woods-Saxon one does the
same in  the region of deformed nuclei.  However, these models are not expected to be adequate
for the description of differential charge radii due to the reasons mentioned below.

    The lack of self-consistency effects and the interaction between protons and neutrons
will affect the description of differential 
charge radii in the mic+mac model. This is because the addition of neutron does not affect the proton 
subsystem in a self-consistent manner on the level of single-particle subshells via the mechanisms 
discussed in Sec.\ \ref{self-const}.  For example, in the Woods-Saxon potential it affects the  total 
radius $R$ of nucleus  only via mass dependence $R=1.2 A^{1/3}$. This means that the occupation  
of the neutron $1i_{11/2}$  and $2g_{9/2}$ subshells in the $N>126$ Pb isotopes will lead to the 
same differential charge radii contrary to the results of self-consistent calculations (see Refs.\ 
\cite{SLR.93,RF.95,PAR.21}). Moreover, there is a lack of self-consistency in the definition  of the 
radial properties of the density distributions in the macroscopic (liquid drop) and microscopic  
(single-particle potential) parts of the mic+mac model.  To our knowledge, this aspect of the problem 
has not been studied in detail.  However, the physical observables similar to  
$\delta \left < r^2 \right>_p^{N,N'}$, namely, relative charge quadrupole 
moments of superdeformed  bands are affected by the lack of self-consistency between microscopic and 
macroscopic parts (see Ref.\ \cite{KRA.98}).

    Although the spherical shell model takes into account the proton-neutron interaction it 
suffers  from the introduction of the core. As a consequence, the pull provided by extra neutron(s) 
on the proton single-particle states forming the core is ignored and this affects drastically the
calculated charge radii of the nuclei with valence nucleons outside the core. This introduces 
uncontrollable errors in the calculations of differential charge radii and thus severely limits the 
applicability of spherical shell model to the description of this observable. Few existing calculations 
of differential charge radii in spherical shell model (see Refs.\ \cite{CLMNV.01,BMKYPCS.21}) 
suffer from this problem. For example, they cannot reproduce the kink in the differential charge 
radii of the Sn isotopes at $N=82$ and Pb isotopes at $N=126$ \cite{BMKYPCS.21}. 
This problem can be rectified by employing no-core shell model but because of numerical reasons 
such models are  applicable only to light nuclei \cite{NQSB.09,BNV.09}.

\subsection{Potential mechanisms affecting odd-even staggering (OES) in charge radii}
%%%%%%%%%%%%%%%%%%%%%%%%%%%%%%%%%%%%%

    Several mechanisms of regular and inverted OES in charge radii 
have been discussed and reviewed in Sec.\ IX of Ref.\ \cite{PAR.21}.  These 
include shape coexistence leading to deformation staggering in even and 
odd-$A$ nuclei (see Sec. IXA of Ref.\ \cite{PAR.21}), pairing correlations 
(see Sec. IXB of Ref.\ \cite{PAR.21}), particle-vibration coupling (PVC) in 
odd-$A$ nuclei (see Sec. IXC of Ref.\ \cite{PAR.21}) and some other 
mechanisms (see Sec. IXD of Ref.\ \cite{PAR.21}). The PVC mechanism is 
responsible for a substantial fragmentation of the wave function of the ground 
states in spherical odd-$A$ nuclei (see Sec. IXC of Ref.\ \cite{PAR.21}) and 
rearrangement of the energies of the predominantly single-particle states 
(see Ref.\ \cite{Pb-Hg-charge-radii-PRL.21}).

   Here we extend the discussion presented in Sec.\ IX of Ref.\ \cite{PAR.21} 
and outline other potential mechanisms which could affect OES in charge radii.
This discussion is based on the fact that it is the single-particle content of 
unpaired neutron states in odd-$N$ nuclei which defines the pull on proton 
densities (see present paper and Refs.\ \cite{GSR.13,PAR.21}). If the 
deformation changes between neighbouring even-even and odd-$A$
nuclei are small, this is dominant mechanism defining OES in charge
radii.
 
   If the structure of extra neutron states does not change with their sequential 
addition, the differential charge radii show a linear evolution as a function of
neutron number (see discussion of Fig. 4 in Ref.\ \cite{PAR.21}) and no OES
in charge radii is present. However, if the structure of the neutron state in
odd-$N$ nuclei is affected by the interaction(s) or effect(s) not present in its 
even-even neighbors (such as blocking effect for pairing
\cite{FTTZ.94,FTTZ.00,RN.17}) and/or particle-vibration coupling 
\cite{Pb-Hg-charge-radii-PRL.21,PAR.21}),  this leads to OES in charge
radii.
 
 The transition to deformed nuclei opens additional channels which are not
present in spherical ones but which can affect OES in charge radii. These 
are the Coriolis interaction in odd and odd-odd nuclei and residual interaction 
of unpaired proton and neutron in odd-odd nuclei: each of them affects the 
structure of the wave function in odd-$N$ nuclei as compared with the ones
of their even $N\pm 1$ neighbors and this can  contribute to OES in 
charge radii. To our knowledge these channels have not been indicated
in literature as possible sources of OES in charge radii because they are
neglected in the DFT calculations. 
  
 The admixtures of the vibrational phonons to the structure of the ground 
states  in deformed rare-earth and actinide odd-mass nuclei  is  relatively small (see 
Ref.\ \cite{GISF.73,ABNSW.88,SSMJ.15}) especially when compared with spherical 
nuclei (see Refs.\ \cite{LR.06,LA.11,AL.15}). Thus, their role in building OES of charge 
radii is expected to be reduced as compared with the one in spherical nuclei (see 
Refs.\ \cite{Pb-Hg-charge-radii-PRL.21,PAR.21}).  On the other hand, the Coriolis
interaction can be active in the ground states of  deformed odd-$N$ nuclei but it is 
absent in the ground states of  even-even nuclei (see Refs.\ 
\cite{JSSJ.90,NilRag-book}). This interaction leads to a $\Delta K=1$ \
mixing\footnote{Here $K$ is the projection of single-particle angular momentum on 
the axis of symmetry of deformed nucleus.} of different states both 
in odd (see Refs.\ \cite{JSSJ.90}) and odd-odd (see  Refs.\ \cite{BPO.76,NKSSN.94}) 
nuclei.  If the wavefunction of the ground state of the nucleus with odd-$N$ 
is strongly affected by Coriolis mixing, this can affect OES of charge radii.      
     
 The situation in odd-$Z$ isotopic chains (see Fig.\ 27 in Ref.\ \cite{PAR.21})
which includes odd-odd nuclei becomes even more complicated for theoretical
interpretation since in addition to Coriolis interaction there is the possibility
for the $\Delta K=0$ mixing due to residual interaction of unpaired proton and 
neutron in odd-odd nuclei \cite{BPO.76,NKSSN.94}. Note that such mixing
is typically calculated in the rotor+particle models (PRM) and it is neglected
in the state-of-the-art DFT applications. In some cases the $\Delta K=0$ 
mixing can be strong so it can potentially have significant impact on the structure
 of the wave function of odd neutron which generates the pull on proton densities.  
 
   Let us consider the isotopic chain of the Eu isotopes 
which is one of the best examples of inverted OES in charge radii
(see Fig. 27 in Ref.\ \cite{PAR.21}). As discussed in 
Sec.\  IX of Ref.\ \cite{PAR.21} the inversion of OES at $N\approx 88$ is most likely
triggered by the transition from spherical or quasispherical nuclei to deformed ones 
with increasing neutron number. However, such inversion of OES exists also for 
$N=90, 91$ and 92 which covers deformed nuclei (see Fig. 27(f) in Ref.\ \cite{PAR.21}).
The experimental observations (Ref.\ \cite{154Eu}) and PRM calculations  (Ref.\ \cite{GATB.87}) indicate 
strong $\Delta K=0$ mixing  for the $K^{\pi}= 1^{\pm}, 3^{\pm}$ and $4^{\pm}$ 
bandheads in the $^{154}$Eu nucleus including $K^{\pi}=3^-$ ground state with 
the $\pi 5/2[413] -\nu11/2[505]$  structure. Such mixing is observed and calculated 
in two-quasiparticle configurations of $^{156}$Eu \cite{156Eu} but ground state 
of this nucleus seems not to be affected by it.  These features may lead to the inversion 
of OES around $N=91$ and its absence above $N=92$. In reality, at the neutron 
numbers of interest the neighboring $Z=62$ Sm isotopes show regular OES in 
charge radii (see Fig. 26(l) in Ref.\ \cite{PAR.21}) which suggests that the inversion
of OES in charge radii of the Eu ($Z=63$) isotopes are most likely due to the properties
of odd-odd nuclei. The detailed DFT calculations with residual interaction 
of unpaired proton and neutron in odd-odd nuclei are needed to confirm or reject such 
an interpretation. However, such investigation goes beyond the scope of the present 
study.

%%%%%%%%%%%%%%%%%%%%%%%%%%%%%%%%
\section{Conclusions}
\label{Concl}
%%%%%%%%%%%%%%%%%%%%%%%%%%%%%%%%  
 
    The self-consistency and proton-neutron interaction effects
in the buildup of differential charge radii have been considered 
by comparing two configurations of the $^{218}$Pb nucleus,
generated by the occupation of the neutron $1i_{11/2}$ and 
$2g_{9/2}$ subshells, with the ground state configuration in 
$^{208}$Pb. The main contribution to differential charge radii
has a single-particle origin and comes from the interaction of added 
neutron(s) and the protons forming the $Z=82$ proton core.  This 
interaction depends on the overlaps of the 
proton and neutron wave functions and leads to a redistribution 
of single-particle density of occupied proton states which in turn
modifies the charge radii.  In contrast, self-consistency effects affecting 
the shape of proton potential, proton densities and the energies of the 
single-particle states in the proton potential provide only a small 
contribution. Note that  the buildup of differential charge radii between
two isotopes is also a collective phenomenon since all occupied
proton single-particle states contribute to it.  Although these results 
were obtained for the Pb isotopes, they are general and applicable to 
any isotopic chain.

   This combination of single-particle and collective aspects in the
building of differential charge radii limits the applicability of different
classes of the models to the description of this physical observable.
The models which ignore self-consistency effects and proton-neutron
interaction (such as microscopic+macroscopic model) cannot describe
this physical observable. The models which ignore collective aspect
of the problem by introducing the core (such as spherical shell model)
introduce uncontrollable errors in the description of differential charge
radii.

   The pull provided by an extra neutron on proton states depends on the 
structure of the state occupied by it. For deformed even-$Z$ nuclei, the wave 
function of the ground states in odd-$N$ nuclei is more complicated 
than that of the ground states in neighboring even-even nuclei because of the 
presence of the Coriolis interaction. The addition of proton (odd-$Z$ 
nuclei) leads to a further complication of the wave function since Coriolis
interaction is present in all nuclei of isotopic chain under study and, in
addition,  the structure of odd-odd nuclei is affected by the residual interaction 
between unpaired proton and neutron.  The analysis indicates than 
both the Coriolis interaction in odd and odd-odd nuclei and the residual
interaction between unpaired proton and neutron in odd-odd nuclei
could affect the odd-even staggering in charge radii if their impact
on the wave function of the ground state of these nuclei is appreciable. 
Note that these interactions are usually neglected in the DFT calculations.

%%%%%%%%%%%%%%%%%%%
\section{ACKNOWLEDGMENTS}
%%%%%%%%%%%%%%%%%%%

 This material is based upon work supported by the U.S. Department of Energy,  
Office of Science, Office of Nuclear Physics under Award No. DE-SC0013037.

\bibliography{references-40-radii-self-consistency-2023}

\end{document}